\documentstyle[aps,epsf,multicol]{revtex}

\newcommand{\extraspace}{\addtolength{\abovedisplayskip}{2mm} 
                        \addtolength{\belowdisplayskip}{2mm} 
                        \addtolength{\abovedisplayshortskip}{2mm} 
                        \addtolength{\belowdisplayshortskip}{2mm}} 
\newcommand{\be}{\begin{equation}\extraspace} 
\newcommand{\ee}{\end{equation}} 
\newcommand{\bea}{\begin{eqnarray}\extraspace} 
\newcommand{\eea}{\end{eqnarray}}

\begin{document} 

\widetext 
\title{Comment on the paper ``The universal chiral partition 
       function for exclusion statistics''}
\author{Kareljan Schoutens}  
\address{ 
     Institute for Theoretical Physics,
     Valckenierstraat 65, 1018 XE Amsterdam, 
     THE NETHERLANDS} 
\date{August 27, 1998 --- revised July 8, 1999}
\maketitle
\begin{abstract} 
I comment on the paper hep-th/9808013 by A.~Berkovich
and B.M.~McCoy.

\end{abstract}


\vskip 6mm

\begin{multicols}{2}

The original version of this Comment was a reaction
to the first version of the paper \cite{BM}. My main
criticism was that in this paper, the authors did
not properly acknowledge existing results in the literature, 
where character formulas based on fractional exclusion 
statistics had been displayed explicitly (see e.g.
\cite{Hi,ES}). In addition, I observed that, despite their 
claim of presenting a `Universal Chiral
Partition Function', the authors did not investigate 
the connection between their formulas and recent,
explicit results on the exclusion statistics of
quasi-particles in conformal field theory. In particular,
they did not recognize the connection with the 
concept of `non-abelian exclusion statistics', which was 
first proposed in the context of edge excitations over 
non-abelian quantum Hall states \cite{Sc}. They also
failed to mention that in important examples, which include 
several higher-rank, level-1 WZW models, we see natural 
quasi-particle character formulas that are not of the 
`Universal Chiral Partition Function' form.
  
After the publication of the original Comment, the authors 
have revised their manuscript and added a number of references.
Furthermore, my remarks concerning non-abelian exclusion 
statistics and more general character formulas have been 
substantiated in a number of recent papers \cite{BS,GS,BCR}. 
In view of these developments, I have replaced the original 
Comment by the present brief statement.

\end{multicols}

\end{document}